\documentclass[11pt,a4paper]{article}
\usepackage{cite}
\usepackage{cutwin}
\usepackage[utf8]{inputenc}
\usepackage{amsmath,amsfonts,amscd,amssymb,mathrsfs}
\DeclareMathOperator{\Tr}{Tr}

\usepackage[T1]{fontenc}
\usepackage[english]{babel}
\usepackage{epsfig}
\usepackage{changebar}
\usepackage{lmodern} 
\usepackage[all]{xy}
\usepackage{mathtools}
\usepackage{MnSymbol}
\usepackage{fancybox}
\usepackage{wrapfig}
\usepackage{cite}
\usepackage{hyperref}
\usepackage{fancyhdr}
\usepackage{vmargin} 
\usepackage[nottoc, notlof, notlot]{tocbibind}
\usepackage{graphicx}
\usepackage{caption}
\usepackage{subcaption}
\usepackage{titlesec}
\usepackage{color}
\usepackage{stmaryrd}
\usepackage{slashed}
\usepackage[most]{tcolorbox}
\usepackage{wrapfig,lipsum,booktabs}
\titleformat{\chapter}[display]   
{\normalfont\huge\bfseries}{\chaptertitlename\ \thechapter}{20pt}{\Huge}   
\titlespacing*{\chapter}{0pt}{-50pt}{40pt}

\titlespacing*{\section}{0pt}{0.2\baselineskip}{\baselineskip}

\DeclareMathOperator*{\tr}{Tr}
\def \k{\kappa}
\DeclareUnicodeCharacter{2212}{-}
\begin{document}

\newcommand\encadremath[1]{\vbox{\hrule\hbox{\vrule\kern8pt
\vbox{\kern8pt \hbox{$\displaystyle #1$}\kern8pt}
\kern8pt\vrule}\hrule}} \def\enca#1{\vbox{\hrule\hbox{
\vrule\kern8pt\vbox{\kern8pt \hbox{$\displaystyle #1$} \kern8pt}
\kern8pt\vrule}\hrule}}

\def\Long{{\color{red} $ \Rightarrow$  Long}
\bigskip
}
\def\red{ \color{red} }
\def\blue{ \color{blue}}


\thispagestyle{empty}

\begin{flushright}
\end{flushright}

\vspace{1cm}
\setcounter{footnote}{0}

\begin{center}

 {\Large\bf  Boundary entropy of integrable perturbed $SU(2)_k$ WZNW}

\vspace{20mm} 

Ivan Kostov,
 Didina Serban and
Dinh-Long Vu \\[7mm]
 
{\it  Institut de Physique Th\'eorique, CNRS-UMR 3681-INP,
 C.E.A.-Saclay, 
 \\
 F-91191 Gif-sur-Yvette, France}, 
 \\[5mm]

\end{center}

\vskip9mm

\vskip18mm

{ \noindent{ We  apply the recently developped  analytical methods for computing the boundary entropy,
or the g-function, in  integrable theories with non-diagonal scattering.
We  consider the particular   case  of  the  current-perturbed $SU(2)_k$  WZNW  model with  boundary and compute the boundary entropy  for a specific boundary condition. The main problem we encounter is that in case of non-diagonal scattering the boundary entropy is  infinite. We show that this infinity can be cured by a subtraction. The difference of the boundary entropies in the UV and in the IR limits is finite, and matches  the  known g-functions for the  unperturbed $SU(2)_k$  WZNW model for even values  of the level. } }
\newpage
\setcounter{footnote}{0}

\section*{Introduction}
Although the Thermodynamic Bethe Ansatz (TBA) \cite{Yang:1968rm, Zamolodchikov:1989cf} technique has been widely applied for integrable periodic systems, less is known about how it could be used to compute boundary related quantities. An important example is the boundary entropy,
which is given by the $O(1)$ correction to the finite temperature partition function of an one dimensional quantum system with open boundary condition
\begin{align}
\log g_a(R)=\frac{1}{2}\lim_{L\to \infty}[\log Z_{aa}(R,L)-\log Z(R,L)].
\end{align}
Here $R$ denotes the inverse temperature, $L$ is the system volume and $a$ is the boundary condition. In the conformal limit $R\to 0$, it is also known as Affleck and Ludwig's g-function \cite{Affleck:1991tk}. By abuse of language we call it g-function even for off-critical theories.

In a theory with diagonal scattering, an analytic expression for the boundary entropy is known. If we denote the particle types by $n$, the rapidity variable by $u$, the scattering derivatives by $K_{nm}(u,v)\equiv -i\partial_u\log S_{nm}(u,v)$, the reflection derivative corresponding to a boundary condition of type $a$ by $K_n^a(u)\equiv -i\partial_u\log R_n^a(u)$, the pseudo-energies  at inverse temperature $R$ by $Y_n$ then
\begin{gather}
2\log g_a(R)=2\log g_a^\textnormal{trees}(R)+2\log g_a^\textnormal{loops}(R),\label{trees+loops}\\
2\log g_a^\textnormal{trees}(R)=\sum_{n}\int_{-\infty}^\infty\frac{du}{2\pi}[K_{n}^a(u)-K_{nn}(u,-u)-\pi\delta(u)]\log[1+Y_n(u)],\label{trees}\\
2\log g_a^\textnormal{loops}(R)=\log\det\frac{1-\hat{K}^-}{1-\hat{K}^+},\label{loops}
\end{gather}
where the kernels $\hat{K}^\pm$ have support on $\mathbb{R}^+$ and their actions are given by
\begin{gather*}
\hat{K}_{nm}^\pm(F)(u)=\int_0^{+\infty}\frac{dv}{2\pi}K^\pm_{nm}(u,v)f_m(v)F(v),\\
\text{with }\quad K_{nm}^\pm(u,v)\equiv K_{nm}(u,v)-K_{nm}(u,-v),\quad f_m=\frac{Y_m}{1+Y_m}.
\end{gather*}

The derivation of the above formula has a relatively long history. The first attempt to derive it from TBA was done in \cite{LeClair:1995uf} where the authors obtained expression \eqref{trees} as the saddle point approximation. Fluctuations around the saddle point were then computed by Woynarovich in \cite{Woynarovich:2004gc}, yielding the denominator of \eqref{loops}. In \cite{Dorey:2004xk}, the authors took a new approach and considered the low temperature expansion of the partition function. After explicitly working out the first few terms, they conjectured a series expansion for the g-function in the same spirit as a Leclair-Mussardo series \cite{Leclair:1999ys}. In \cite{Pozsgay:2010tv} Pozsgay cast this series into the form \eqref{loops} and interpreted the nominator as a non-trivial functional integration measure around the TBA saddle point, thus complementing the work of Woynarovich. In \cite{Kostov:2018dmi} we re-derived this result by generalizing the approach of \cite{Dorey:2004xk} to arbitrary number of particles. We obtained a graph expansion of the partition function with help of the matrix-tree theorem. The boundary-dependent term \eqref{trees} was expressed as a sum over trees while the boundary-independent term \eqref{loops} is a sum over loops.

The analytic expression \eqref{trees+loops} has been verified for a wide class of massive integrable theories with diagonal bulk scattering and boundary reflections \cite{Dorey:2005ak}. For massless theories, only minor modification in the form of the TBA equations is needed \cite{Dorey:2009vg}.

A general formula as \eqref{trees+loops} is not known for theories in which the bulk scattering is not diagonal. Let us first spell out why the methods of \cite{Pozsgay:2010tv} and \cite{Kostov:2018dmi} are not justified in this case. The diagonalization by the Nested Bethe Ansatz technique involves particles of magnonic type, which are auxiliary particles with zero momentum and energy. The functional integration measure, as it is derived
in \cite{Pozsgay:2010tv}, as well as the the summation over multiparticle states in \cite{Kostov:2018dmi} or \cite{Dorey:2004xk} treat the physical
and the auxiliary particles in exactly the same way. This is justified only for states with asymptotically large number of physical particles. For states with finite number of physical particles, which dominate in the IR limit, the solutions of the Bethe Ansatz equations do not obey the string hypothesis and moreover the number of the magnons and the number of the physical particles must respect certain constraint. On the other hand, finding and summing over the exact solutions for the auxiliary magnons is of course a hopeless task.

 In this paper we demonstrate on a concrete example that assuming solutions in the form of Bethe strings and summing over unrestricted number of auxiliary particles nevertheless leads to a meaningful result for the boundary entropy, up to an infinite constant which can be subtracted. The subtraction is done by normalizing the g-function \eqref{trees+loops} by its zero temperature value.
 
The appearance of this infinite piece has an obvious explanation. When the theory is massive in the bulk, the boundary entropy must vanish at zero temperature. If the bulk scattering is diagonal, this condition  is automatically satisfied by the expression \eqref{trees+loops} as all pseudo-energies vanish in this limit. For non-diagonal bulk scattering however, magnonic particles decouple from the physical ones at zero temperature and retain non-zero pseudo-energies. In the formalism of \cite{Kostov:2018dmi}, our normalization amounts to subtracting the contribution from unphysical graphs made of these auxiliary particle. We denote in this paper $g_{\text{IR}}\equiv g(\infty),\; g_{\text{UV}}\equiv g(0)$
\begin{align}
g(R)^\text{ren}= \frac{g(R)}{g_{\text{IR}}}.\label{proposition}
\end{align}
As a test for this proposal, we show that it is possible to match the ratio $g_{\text{UV}}/g_{\text{IR}}$ with a conformal g-function under certain assumptions.

The theories under study are the current-perturbed $SU(2)$ Wess-Zumino-Novikov-Witten (WZNW) model at positive integer levels $k$. The bulk scattering of these theories are not diagonal, as each particle carries quantum numbers that can be nontrivially exchanged during collisions. With the Nested Bethe Ansatz technique, one can trade the nondiagonal scattering for a diagonal one with extra magnonic particles: $SU(2)$ magnon and kink magnon. In the thermodynamics limit these particles can form bound states which are strings of evenly distributed rapidities on the complex plane. In particular $SU(2)$ magnons can form strings of arbitrary lengths, effectively leading to an infinite number of particles in the TBA formalism. In the derivation of the TBA free energy one ignores that the above is true only for asymptotically large number of physical particles. The price to pay, as we will show later, is that both expressions \eqref{trees},\eqref{loops} are logarithmic divergent in the IR and UV limit \footnote{Note however that the bulk free energy computed in this way does not diverge and reproduces the correct
degeneracy of levels in the IR limit \cite{Ahn:2011xq}.}.

We regularize these divergencies by introducing a twist or equivalently a chemical potential to the TBA equations. The chemical potential makes the sum over the auxiliary magnons finite. The twist/chemical potential is added to the TBA equations for the sole purpose of regularizing the g-function and we do not discuss its effects on the UV limit of the theory. The only change induced by this modification in the formulae \eqref{trees} and \eqref{loops} is the asymptotic values of Y-functions. We are then able to express the IR and UV g-function as functions of the twist parameter and evaluate their ratio in the untwisted limit. For a specific choice of boundary condition it is given by  
\begin{align}
\bigg(\frac{g_{\text{UV}}}{g_{\text{IR}}}\bigg)^2=\sqrt{\frac{2}{k+2}}\times\frac{1}{\sin\frac{\pi}{k+2}},\label{final-result}
\end{align}
which coincides with a Cardy g-function \eqref{CFT-entropy}, namely $g_{k/2}$ for even $k$. Equation \eqref{final-result} is the main result of this paper.

The paper is organized as follows. In section 1 we present the basic features of $SU(2)$ WZNW CFT at level $k$ with emphasis on its Cardy g-functions. We also show how the boundary entropy of a massive perturbation of this CFT flows to its UV value when the temperature is sent to infinity. In section 2 we introduce the current perturbation of this CFT and its TBA equations. We show how various quantities can be extracted from solutions of the TBA equations in the UV or IR limit. We conjecture a specific set of diagonal reflection factors in section 3, fixing the values of $K_n^a(u)$ in equation \eqref{trees}. With the data from section 2 and 3, we show in section 4 that the expressions \eqref{trees} and \eqref{loops} diverge in the UV and IR limit. We then show using  twisted TBA equations that these divergencies can be regularized, leading to \eqref{final-result} as the final result.

Although a general method for finding the g-function of a theory with non-diagonal bulk scattering is still missing, there are models with particular features that allow this quantity to be extracted via case-dependent techniques. In \cite{Dorey:2010ub}, the g-functions of perturbed unitary minimal models   were studied using the roaming trajectory of the staircase model \cite{Zamolodchikov:1991pc}. The latter is a theory with diagonal bulk scattering that depends on a free parameter. This parameter can be tuned with temperature to form a plateau RG flow with successive unitary minimal models on its steps. Recently Pozsgay \cite{2018arXiv180409992P} computed the spin chain analogue of g-function for the XXZ spin chain using quantum transfer matrix and independent results on integrable overlaps. The TBA of this spin chain also involves an infinite number of magnon strings. It could be observed from the results of  \cite{Dorey:2010ub} and \cite{2018arXiv180409992P} that the Fredholm determinant structure \eqref{loops} of the g-function is still relevent for theories with non-diagonal bulk scattering.
\section{Setup}
\subsection{$SU(2)_k$ WZNW CFT and its boundary states}
\label{sec:2}
The Wess-Zumino-Novikov-Witten (WZNW) model for a semisimple group G is defined by the action
\begin{align*}
S_{\text{WZNW}}=\frac{1}{4\lambda^2}\int_{S^2} d^2x\tr\partial_\mu g\partial^\mu g^{-1}+k\Gamma,
\end{align*}
where the Wess-Zumino term $\Gamma$ is 
\begin{align*}
\Gamma=\frac{1}{24\pi}\int_{B^2} d^3X\epsilon^{ijk}\tr \tilde{g}^{-1}\partial_i \tilde{g}\tilde{g}^{-1}\partial_j \tilde{g}\tilde{g}^{-1}\partial_k \tilde{g}.
\end{align*}
Here $g$ is a map from the two-sphere to $G$ and $\tilde{g}$ is its extension from the corresponding two-ball to the same group. Such an extension comes with an ambiguity of topological origin, leading to integer values of $k$.

At $\lambda^2=4\pi/k$ the global $G\times G$ symmetry is enhanced to a local $G(z)\times G(\bar{z})$ symmetry with two currents $J(z)=\partial_z g g^{-1}$ and $\bar{J}(\bar{z})=g^{-1}\partial_{\bar{z}}g$ separately conserved. These currents satisfy the current algebra $G_k$ while their bilinear satisfies the Virasoro algebra. The latter implies in particular conformal invariance and we refer to the theory at this coupling as the WZNW CFT of $G$ at level $k$.

In the following we consider the case $G=SU(2)$. The left(right) moving sector of this theory consists of $k+1$ irreducible representations $\mathcal{V}_\lambda$ of $SU(2)_k$ corresponding to its $k+1$ integrable  weights. The characters $\chi_\lambda=q^{-c/24}\tr_{\mathcal{V}_\lambda}q^{L_0}$ transform to one another under the modular transformation $\tau\to -1/\tau$. This transformation is encoded in the modular S-matrix of the theory
\begin{align*}
\chi_\lambda(q)=\sum_{\eta=0}^kS_{\lambda,\eta}\chi_\eta(\tilde{q}),\quad q\equiv e^{2i\pi\tau},\tilde{q}\equiv e^{-2\pi i/\tau}.
\end{align*}
It is explicitly given by
\begin{align}
S_{\lambda,\eta}=\sqrt{\frac{2}{k+2}}\sin\bigg[\frac{\pi(\lambda+1)(\eta+1)}{k+2}\bigg],\quad 0\leq \lambda,\eta\leq k,
\label{SU(2)-WZW-level-n-modular}
\end{align}
which is a real, symmetric matrix that satisfies $S^2=\mathbf{1}$. The central charge and conformal dimesnsions are obtained from the Sugawara construction of the energy-momentum tensor
\begin{align}
c=\frac{3k}{k+2},\quad h_\lambda=\frac{\lambda(\lambda+2)}{4(k+2)}.\label{WZW-n}
\end{align}
The fusion coefficients $\mathcal{N}_{\lambda,\eta}^\kappa$ denote how many times the field $\phi_\kappa$ appears in the operator product expansion of $\phi_\lambda$ and $\phi_\eta$. They satisfy the Verlinde formula
\begin{align}
\mathcal{N}_{\lambda,\eta}^\kappa=\sum_{\zeta}\frac{S_{\lambda,\zeta}S_{\eta,\zeta}S_{\zeta,\kappa}}{S_{0,\zeta}}.\label{Verlinde-formula}
\end{align}
Above is our quick summary of $SU(2)_k$ WZNW CFT data. Consider now this CFT on manifolds with boundaries. Two geometries are relevant for our discussion. First let us consider the upper half complex plane. The continuity condition through the real axis requires that the underlying symmetry involves only one copy of the algebra instead of two. A particular set of boundary states which are invariant under the Virasoro algebra as well as the affine $SU(2)$ Lie algebra is called Ishibashi states \cite{Ishibashi:1988kg}. These states are in one-to-one correspondence with bulk primaries and are denoted as $|\lambda\rrangle$
\begin{align}
(L_n-\tilde{L}_{-n})|\lambda\rrangle =(J^a_n+\tilde{J}^a_{-n})|\lambda\rrangle=0.
\end{align}
Other boundary states are obtained from linear combination of Ishibashi states
\begin{align}
|a\rangle=\sum_{\lambda}|\lambda\rrangle\llangle \lambda|a\rangle.\label{linear-combination-Ishibashi}
\end{align}
So, on the upper half complex plane, one is quite free to choose the boundary state. The situation is greatly different if the theory is restricted on an annulus. In this geometry let us consider two boundary states $|a\rangle$ and $|b\rangle$ of the form \eqref{linear-combination-Ishibashi} on its sides. Denote by $q=\exp(-\pi R/L)$ the modular parameter of this annulus. On one hand one can quantize this theory according to the Hamiltonian $H_{ab}$ with $a,b$ as boundary conditions
\begin{align}
Z_{ab}=\sum_{\lambda}n_{a,b}^\lambda\chi_\lambda(q)=\sum_{\lambda}n_{a,b}^\lambda\sum_\eta S_{\lambda,\eta}\chi_\eta(\tilde{q}),\quad \tilde{q}=\exp(-4\pi L/R),\label{partition-function-1}
\end{align}
where the non-negative integers $n_{a,b}^\lambda$ denote the number of copies of $\mathcal{V}_\lambda$ in the spectrum of $H_{ab}$. On the other hand one can consider the theory as evolving between two states $\langle a|$ and $|b\rangle$. The periodic Hamiltonian can be written in terms of Virasoro generators via a conformal mapping, leading to
\begin{align}
Z_{ab}=\langle a|\tilde{q}^{\frac{1}{2}(L_0+\bar{L}_0-c/12)}|b\rangle=\sum_{\eta}\langle a|\eta\rrangle\llangle \eta|b\rangle\chi_\eta(\tilde{q}).\label{partition-function-2}
\end{align}
\begin{figure}[h]
\centering
\includegraphics[width=10cm]{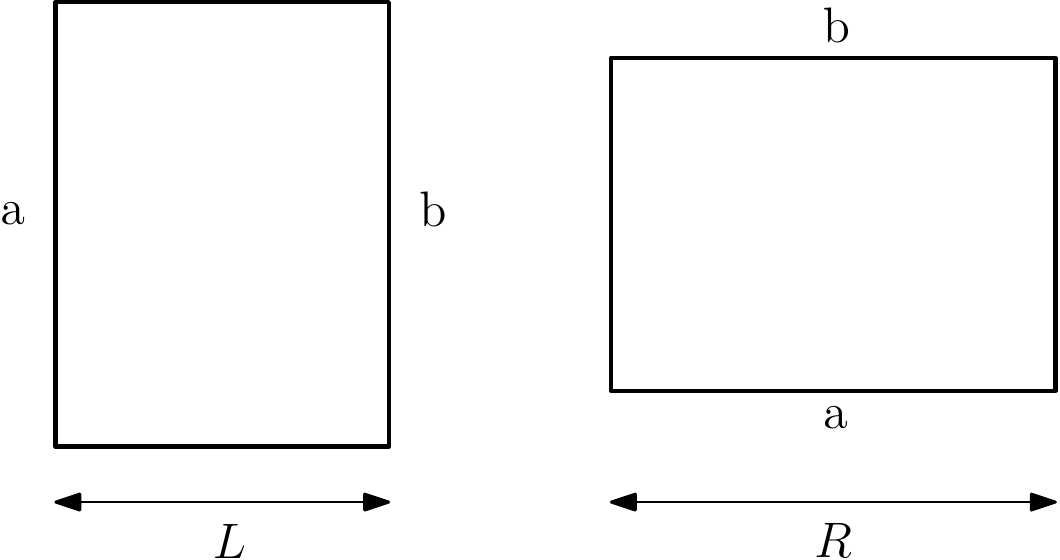}
\caption{Modular invariance of the annulus partition function.}
\end{figure}

For the theory under consideration, each representation $\lambda$ of the extended algebra appears only once in the spectrum. By identifying the two expressions \eqref{partition-function-1} and \eqref{partition-function-2} we obtain the following relation 
\begin{align}
\sum_\lambda n_{a,b}^\lambda S_{\lambda,\eta}=\langle a|\eta\rrangle \llangle \eta|b\rangle\Leftrightarrow n_{a,b}^\lambda=\sum_\eta S_{\eta,\lambda}\langle a|\eta\rrangle \llangle \eta|b\rangle,\label{Cardy-equation}
\end{align}
where we have used the fact that $S$ is real symmetric and $S^2=\mathbf{1}$. Relation \eqref{Cardy-equation} is referred to as Cardy equation \cite{Cardy:1989ir}, which sets the constraint on admissible boundary states on an annulus. A particular solution is given by
\begin{align*}
\langle a|\lambda\rrangle=\frac{S_{a,\lambda}}{\sqrt{S_{0,\lambda}}},\quad n=\mathcal{N},
\end{align*}
where the Cardy equation becomes the Verlinde formula \eqref{Verlinde-formula}. We refer to these boundary states as Cardy states and denote them by $|C_a\rangle,\; a=\overline{0,k}$.

Due to the difficulty in defining the Wess-Zumino term for a surface with boundary, our discussion of g-function should be taken at the operatorial level. We also stress that the  boundary states under consideration are invariant under the extended algebra. For boundary states which are only conformal invariant, see \cite{Gaberdiel:2001xm}. 
\subsection{Off-critical g-function}
Let us now assume that there is an 1+1 dimensional integrable massive quantum field theory which admits $SU(2)_k$ WZNW CFT as its UV fixed point. We consider such a theory on a cylinder of length $L$ and radius $R$ which plays the role of periodic Euclidean time or equivalently, inverse temperature. We further assume that we can define the boundary conditions in such a way that integrability is conserved \footnote{At the quantum level, this is guaranteed if one can find a solution of the boundary Yang-Baxter equation. See section 3 for further discussion.}. 

One would expect from integrability that, at arbitrary temperature, it is possible to compute the bulk free energy and boundary entropy densities of the theory
\begin{align*}
Z_{ab}(R,L)=\exp[-LRf(R)]\times g_a(R) g_b(R).
\end{align*}
We also assume that in the conformal limit $R\to 0$ the two integrable boundary conditions can be identified with some CFT boundary sates $|a\rangle$ and $|b\rangle$ \footnote{This hypothesis is usually satisfied for integrable boundary conditions, see for instance \cite{Dorey:1999cj, Dorey:2005ak,Dorey:2010ub}}. Then in this limit the modular parameter $q$ tends to one and the contribution of vacuum state dominates other terms in the partition function \eqref{partition-function-1}
\begin{align*}
\lim _{R\to 0}Z_{ab}(R,L)=\chi_0(\tilde{q})\sum_{\lambda}n_{a,b}^\lambda S_{\lambda,0}.
\end{align*} 
Therefore the bulk free energy becomes proportional to the CFT central charge
\begin{align}
\lim_{R\to 0}R^2f(R)=-\frac{\pi c}{6} \label{CFT-energy}
\end{align} 
while the boundary contribution to the partition function is given by the sum
\begin{align*}
g_a(0)g_b(0)= \sum_{\lambda}n_{a,b}^\lambda S_{\lambda,0}.
\end{align*}
Apply now the Cardy equation \eqref{Cardy-equation}, one can identify the contribution of each boundary with the corresponding overlap with the Ishibashi state $|0\rrangle$. In particular, if the boundary state  happens to be a Cardy state, we expect the boundary entropy to flow to 
\begin{align}
g_a(0)=\frac{S_{a,0}}{\sqrt{S_{0,0}}}\label{CFT-entropy}
\end{align}
in the UV limit. 

It is the purpose of this paper to give the exact expression for the boundary entropy $g_a$ at arbitrary temperature for a particular perturbation of $SU(2)_k$ WZNW CFT and to match its value with some Cardy g-function in the UV limit. First, we remind how to compute the bulk free energy $f(R)$ using the Thermodynamic Bethe ansatz technique. In particular we will verify the limit \eqref{CFT-energy}.
\section{ TBA equations for current-perturbed $SU(2)_k$ WZNW}
\label{sec:3}
The perturbation we are going to consider belongs to a larger family of perturbations of diagonal coset CFT's $G_k\times G_l/G_{k+l}$ where $G$ is simply-laced. It was first shown in \cite{AHN1990409} that it is possible to perturb this CFT while still preserving part of its symmetry. The perturbing operator is the branching between two scalar representations and the adjoint representation in the Goddard-Kent-Olive (GKO) construction. Moreover, for negative sign of the perturbing parameter, this perturbation leads to a massive field theory. In the same paper, a factorized scattering matrix for this field theory in the particular case of $G=SU(2)$ was proposed 
\begin{align}
\mathcal{S}=\mathcal{S}_{[k]}^\text{RSG}\otimes\mathcal{S}^\text{RSG}_{[l]},\label{S-matrix-k-l}
\end{align} 
where RSG stands for restricted sine-Gordon. Each factor in this tensor product is a S-matrix of the sine-Gordon theory at coupling $\beta^2/8\pi=(k+2)/(k+3)$ and $(l+2)/(l+3)$ respectively. The idea is to rely on the quantum group symmetry of the sine-Gordon S-matrix when the deformed parameter is a root of unity ($q=-\exp[-i\pi/(k+2)]$) to restrict the original multisoliton Hilbert space to direct sum of irreducible representations of this quantum group. As a result the infinite set of vacua of the sine-Gordon theory is truncated to $k+1$ $(l+1)$ vacua and a particle of the restricted theory is defined to be a kink interpolating adjacent vacua. This case was further studied in \cite{Zamolodchikov:1991vg} where a system of TBA equations was conjectured and shown to yield the correct central charge of the unperturbed CFT in the UV limit. It should be noted that the TBA system in \cite{Zamolodchikov:1991vg} was not based on the scattering reported in \cite{AHN1990409} but was instead taken as generalization of several known cases. A derivation of TBA equations for $G=SU(N)$ from the basis of scattering data was later carried out in \cite{HOLLOWOOD199443}. The extension to other simple Lie algebras was done in \cite{Babichenko:2003rf}. See also \cite{Bazhanov:1989yk} for a lattice realization.

In this paper we are interested in the limit $l\to \infty$ and $G=SU(2)$ where the coset CFT reduces to $SU(2)_k$ WZNW  and the branching operator becomes its Kac-Moody current. Moreover the second factor in the S-matrix \eqref{S-matrix-k-l} becomes the S-matrix of the chiral $SU(2)$ Gross-Neveu model.  We first study this theory independently.
\subsection{Chiral $SU(2)$ Gross-Neveu model and its TBA equations}
The chiral $SU(N)$ Gross-Neveu model (also known as the massive $SU(N)$ Thirring model) is conventionally defined via an action involving $N$ Dirac fermions
\begin{align}
\mathcal{L}=\int dx^2\bar{\psi}_ai\slashed{\partial}\psi_a+g[(\bar{\psi}_a\psi_a)^2-(\bar{\psi}_a\gamma^5\psi_a)^2],\quad a=\overline{1,N}.\label{GN-action}
\end{align}
Using the technique of nonabelian bosonization \cite{Witten:1983ar}, this action was shown to be equivalent to the action of current-perturbed $U(N)_1$ WZNW theory \cite{DiVecchia:1984df},\cite{Knizhnik:1984nr}. The $U(1)$ center is identified with a massless boson that decouples from the rest of the spectrum, which we shall refer to as the chiral $SU(N)$ Gross-Neveu model. For $N=2$ it is also equivalent to sine-Gordon theory at $\beta^2=8\pi$, thus explaining the $l\to\infty$ limit in \eqref{S-matrix-k-l}.

Each particle of this theory is in one-to-one correspondence with fundamental representations of the $SU(N)$ group. The particle corresponding to the Young tableau of one column and $a$ rows is a bound state of $a$ vector particles and has mass
\begin{align*}
m_a=m\frac{\sin \pi a/N}{\sin a/N},\quad a=\overline{1,N-1}.
\end{align*} 
The scattering matrix between vector particles can be elegantly derived from Yang-Baxter equation, unitarity and crossing symmetry along with bound state structure \cite{Fendley}. To this end one obtains the so-called minimal S matrix
\begin{align}
\mathcal{S}_{\text{VV}}^{\text{SU(N)}}(\theta)=\frac{\sinh(\theta/2+ i\pi/N)}{\sinh(\theta/2-i\pi/N)}\times\frac{\Gamma(1-\frac{\theta}{2\pi i})\Gamma(1/N+\frac{\theta}{2\pi i})}{\Gamma(1+\frac{\theta}{2\pi i})\Gamma(1/N-\frac{\theta}{2\pi i})}\times\bigg(\mathcal{P}_\text{S}+\frac{\theta+2\pi i/N}{\theta-2\pi i/N}\mathcal{P}_\text{A}\bigg),\label{SU(N)-GN-Scattering-matrix}
\end{align}
where $\mathcal{P}_{\text{S}}$ and $\mathcal{P}_{\text{A}}$ are projections to the symmetric and antisymmetric representations that appear in the tensor product of two vector representations. This matrix can be used as elementary block to construct the scattering matrix between any particles of the theory. By simplifying the product of scalar factors and writing the matrix part in terms of the permutation operator, we can rewrite \eqref{SU(N)-GN-Scattering-matrix} as
\begin{align}
\mathcal{S}^\text{SU(N)}_\text{VV}(\theta)=\frac{N\theta-2\pi i\mathcal{P}}{N\theta-2\pi i}S_0^{\text{SU(N)}}(\theta),
\end{align}
with
\begin{align}
S_0^{\text{SU(N)}}(\theta)=-\frac{\Gamma(1-\theta/2\pi i)}{\Gamma(1+\theta/2\pi i)}\frac{\Gamma(1-1/N+\theta/2\pi i)}{\Gamma(1-1/N-\theta/2\pi i)}.
\end{align}

For $N=2$ \footnote{in this case there is no bound state, the CDD factor doesn't introduce any extra pole and is simply $-1$}, there is only one particle in the spectrum so the vector-vector scattering is all we need
\begin{align}
\mathcal{S}^{\text{SU(2)}}(\theta)=\frac{\theta-\pi i\mathcal{P}}{\theta-\pi i}S_0^{\text{SU(2)}},\quad S_0^{\text{SU(2)}}(\theta)=-\frac{\Gamma(1-\theta/2\pi i)}{\Gamma(1+\theta/2\pi i)}\frac{\Gamma(1/2+\theta/2\pi i)}{\Gamma(1/2-\theta/2\pi i)}.\label{scalar-part}
\end{align}
With the scattering matrix at hand, one can now quantize a multiparticle state on a circle of length $L$ which is large compared to the inverse mass scale. The periodic condition imposed on the wave function $|\Psi\rangle =|\theta_1,\theta_2,...,\theta_N\rangle $ when a particle of rapidity $\theta_j$ is brought around the circle and scatters with other particles reads
\begin{align*}
e^{-ip(\theta_j)L}\Psi=\prod_{k\neq j}\mathcal{S}^{\text{SU(2)}}(\theta_j,\theta_k)\Psi,\quad j=\overline{1,N}.
\end{align*}
Using the fact that the scattering matrix at coinciding rapidity is exactly minus the permutation operator, one can cast the above equation in the following form
\begin{align}
e^{-ip(\theta_j)L}\Psi=-T^\textnormal{SU(2)}(\theta_j)\Psi,\quad j=\overline{1,N}.
\label{transfer-matrix-GN}
\end{align}
where $T(u)=\Tr_{\mathbb{C}_u^2}[\mathcal{S}(u,\theta_1)...\mathcal{S}(u,\theta_N)]$ is the transfer matrix. The advantage of writing Bethe equations in this form is that we can now regard the physical rapidities $\theta's$ as non-dynamical impurities on a spin chain. The argument $u$ of the transfer matrix plays the role of the rapidity of an auxiliary particle living in time direction. The transfer matrix can then be diagonalized using the technique of Algebraic Bethe Ansatz. Its eigenvalues are parametrized by a set of rapidities $u_1,...,u_M$, which, in the spin chain picture, are physical rapidities. Referring to appendix \ref{GN-TBA} for more details, we write here the set of Bethe equations that quantize multiparticle states of chiral $SU(2)$ Gross-Neveu model
\begin{gather}
\begin{aligned}
1&=e^{ip(\theta_j)L}\prod_{k\neq j}^NS_0(\theta_j-\theta_k)\prod_{m=1}^M
\frac{\theta_j-u_m+i\pi/2}{\theta_j-u_m-i\pi/2},\quad j=\overline{1,N}\\
1&=\prod_{j=1}^N\frac{u_k-\theta_j-i\pi/2}{u_k-\theta_j+i\pi/2}\prod_{l\neq k}^M\frac{u_k-u_l+i\pi}{u_k-u_l-i\pi},\quad k=\overline{1,M}
\end{aligned}\label{Bethe-GN}
\end{gather}
We see that in the original picture, the rapdities $u's$ correspond to auxiliary particles with vanishing energy and momentum. By construction, their number should  not exceed half the number of physical rapidities.

In the thermodynamics limit $(L\to \infty)$, the dominant contribution to the partition function comes from solutions of the system \eqref{Bethe-GN} with macroscopic numbers of particles $N$ and $M$. As the number of physical particles $N$ plays the role of spin chain length for the auxiliary particles, in this regime a special family of solutions appears. They are strings of  auxiliary rapidities evenly distributed in distance of $i\pi$ and symmetric with respect to the real axis.

With this string hypothesis, we effectively deal with an infinite number of particle species in thermodynamic limit. We label the physical rapidity by $0$ and the auxiliary rapidities by the length of the corresponding string.  The TBA equations at inverse temperature $R$ read
\begin{align}
\log Y_n(u)+Rm\cosh(u)\delta_{n,0}=\sum_{m=0}^\infty K_{mn}\star\log(1+Y_m)(u),\quad n=\overline{0,\infty},
\end{align}
where the scattering kernels $K_{mn}$ are given in \eqref{kernel-0-n},\eqref{kernel-n-m} and the convolution is normalized as
\begin{align*}
f*g(u)=\int_{-\infty}^{+\infty}\frac{d v}{2\pi}f(u-v)g(v).
\end{align*}
As shown in  appendix \ref{GN-TBA}, this system of TBA equations can  be transformed into a local form called Y-system
\begin{align}
\log \mathcal{Y}_n+RE\delta_{n,0}= s\star[\log(1+\mathcal{Y}_{n-1})+\log(1+\mathcal{Y}_{n+1})],\quad n=\overline{0,\infty}.\label{Y-system-GN}
\end{align}
where $\mathcal{Y}_n=Y_n^{2\delta_{0n}-1}$, $\mathcal{Y}_{-1}=0$ and $s$ is a simple kernel whose Fourier transform is given in \eqref{Fourier-s}. Only the Y-function of the physical rapidity enters in the free energy density 
\begin{gather}
Rf(R)=-m\int_{-\infty}^{+\infty}\frac{d\theta}{2\pi}\cosh(\theta)\log[1+Y_0(\theta)].\label{free-energy-density}
\end{gather}

Two special regimes are of interest. At zero temperature the solutions of TBA equations become constants
\begin{gather}
\mathcal{Y}_n^{\text{IR}}=(n+1)^2-1,\quad n\geq 0.\label{IR-GN-not-twist}
\end{gather}
As we turn on the temperature, a plateau structure starts to develop for each Y-function inside the region  from $-\log[2/(mR)]$ to $\log[2/(mR)]$. Ouside of this region  Y-functions retain their IR values while the tops of the plateaux flatten out at height
\begin{gather}
\mathcal{Y}_n^{\text{UV}}=(n+2)^2-1,\quad n\geq 0\label{UV-GN-not-twist}
\end{gather}
There are two consistency checks for these stationary solutions. In the zero temperature limit one would expect the behavior of a non-interacting gas. In particular, as the physical particle belongs to the vector representation of $SU(2)$, the leading contribution to the free energy should be
\begin{align*}
\lim _{R\to 0}Rf(R)=-2\int_{-\infty}^{+\infty}\frac{d\theta}{2\pi}m\cosh(\theta)e^{-Rm\cosh(u)}.
\end{align*} 
By replacing the leading term of the physical Y-function into the expression \eqref{free-energy-density} we see that this is indeed the case \footnote{for constant functions the kernel $s$ acts as square root. For higher order matching between TBA and Luscher correction, see \cite{Ahn:2011xq}}
\begin{align}
Y_0^{\text{IR}}(u)\approx e^{-Rm\cosh(u)}\sqrt{1+\mathcal{Y}_1^{\text{IR}}}=2e^{-Rm\cosh(u)}.
\end{align}
At the UV, the Casimir energy computed from TBA should match the central charge of the unperturbed CFT, as explained in \eqref{CFT-energy}
\begin{align}
c(R)\equiv -\frac{6R^2f(R)}{\pi}=\frac{3}{\pi^2}\int\limits_{-\infty}^{+\infty} mR\cosh(\theta) \log[1+Y_0(\theta)]d\theta.\label{scaled-free-energy-density}
\end{align}
As the temperature approaching infinity the edges of the plateaux possess temperature-independent structures which satisfy scale invariant TBA equations \cite{Bazhanov:1987zu}. As a result, the above integral can be expressed in terms of Roger dilogarithm function \cite{Klassen:1989ui}
\begin{gather*}
\lim_{R\to 0}c(R)=\sum_{n\geq 0}\text{Li}_\text{R}(\frac{1}{1+\mathcal{Y}_n^{\text{IR}}})-\sum_{n\geq 0}\text{Li}_\text{R}(\frac{1}{1+\mathcal{Y}_n^{\text{UV}}})=1,\\
\text{where}\quad \text{Li}_\text{R}(x)\equiv \frac{6}{\pi^2}[\text{Li}_2(x)+\frac{1}{2}\log(x)\log(1-x)].
\end{gather*}
The particle densities can also be easily computed in the UV limit \cite{Klassen:1990dx}. Let us denote by $D_0=N_0/L$ the density of physical particle and $D_a=N_a/L$ that of string of length a, then
\begin{align*}
\lim_{R\to 0} \pi RD_0(R)=\log(1+Y_0^{\text{UV}}),\quad \lim_{R\to 0} \pi RD_a(R)=\log(1+Y_a^{\text{IR}})-\log(1+Y_a^{\text{UV}}).
\end{align*}
In particular the density of physical particle is exactly twice the total density of auxiliary particles in this limit
\begin{align*}
\lim_{R\to 0} \pi RD_0(R)=\log 4,\quad \lim_{R\to 0}\sum_{a=1}^\infty \pi R a D_a(R)=\log 2.
\end{align*}
For later discussion of g-function, we present here also the  TBA equations in the presence of a chemical potential coupled to the $SU(2)$ symmetry. Denote by $\mu$ the chemical potential of the physical particle. In the spin chain language, $\mu$ can be thought of as the strength of an external magnetic field. The auxiliary particle corresponds to spin flipping and is asigned a chemical potential of $-2\mu$. A string of $n$ auxiliary particles have chemical potential $-2n\mu$. The TBA equations now read
\begin{gather}
\begin{aligned}
\log Y_0(u)&=-Rm\cosh(u)+\mu+\sum_{n=0}^{\infty} K_{n,0}\star\log(1+Y_n)(u),\\
\log  Y_n(u)&=-2n\mu+\sum_{m=0}^\infty K_{m,n}\star\log(1+Y_m)(u),\quad n\geq 1\;.
\end{aligned}
\end{gather}
This inclusion of chemical potential does not affect the structure of Y system \eqref{Y-system-GN}, it does affect however the asymptotic values of Y-functions. Write $2\mu=-\log \kappa$, the new IR and UV values of Y functions are given by
\begin{gather}
1+\mathcal{Y}_n^{\text{IR}}(\k)=[n+1]_{\k}^2,\quad 1+\mathcal{Y}_n^{\text{UV}}(\k)=[n+2]_{\k}^2\; ,\label{IR-UV-GN-twist}
\end{gather}
where the $\k$-quantum numbers are defined as
\begin{align*}
[n]_\k\equiv \frac{1+\k+...+\k^{n-1}}{\k^{(n-1)/2}}.
\end{align*}
We can repeat the above analysis for this twisted theory. At zero temperature, the double degeneracy of up/down spin is lifted 
\begin{align*}
Y_0^{\text{IR}}(u)=e^{-Rm\cosh(u)}\sqrt{1+\mathcal{Y}_1^{\text{IR}}(\k)}=[2]_\k e^{-Rm\cosh(u)}.
\end{align*}
In the UV limit the particle densities are now given by
\begin{align*}
\lim_{R\to 0} \pi RD_0(R,\mu)=2\log (1+\k)-\log \k,\quad \lim_{R\to 0}\sum_{a=1}^\infty \pi R a D_a(R,\mu)=\log (1+\k).
\end{align*}
The scaled free energy density
\begin{align*}
c(R,\mu)\equiv -\frac{6R^2f(R)}{\pi}=\frac{3}{\pi^2}\int\limits_{-\infty}^{+\infty} mR\cosh(\theta) \log[1+Y_0(\theta)]d\theta-\frac{6}{\pi}\sum_{a=0}^\infty\mu_a RD_a(R,\mu)
\end{align*}
where $\mu_0=\mu,\;\mu_n=-2n\mu$ for $n\geq 1$, can again be computed in the UV limit with help of Roger dilogarithm function
\begin{align}
\lim_{R\to 0}c(R,\mu)=1-\frac{6\mu^2}{\pi^2}
\end{align}

It should be stressed that the introduction of a chemical potential in the TBA equations serves only at regularizing the g-function. We do not attempt to establish the link with CFT limit. See \cite{EVANS1995469}, \cite{Fendley:1993zt} for discussions in this direction.
\subsection{Generalization to higher level}
Each particle now carries in addition to the $SU(2)$ quantum number a kink quantum number. Each kink connects two adjacent vacua among $k+1$ vacua of the truncated sine-Gordon Hilbert space. One usually calls $(a,a+1)$ a kink and $(a+1,a)$ an anti-kink. When kinks are scattered, only the middle vacuum can be interchanged. For $k=1$ there are two vacua and the only scattering is between kink $(1,2)$ and anti kink $(2,1)$, which is trivial. The S-matrix for the kink scattering $K_{da}(\theta_1)+K_{ab}(\theta_2)\to K_{dc}(\theta_2)+K_{cb}(\theta_1)$ is the RSOS Boltzmann weight
\begin{align*}
\mathcal{S}_{[k]}^\text{kink}(\theta)\begin{pmatrix}
a&b\\
c&d
\end{pmatrix}(\theta)=&\frac{u(\theta)}{2\pi i}\bigg(\frac{\sinh(\pi a/p)\sinh(\pi c/p)}{\sinh \pi d/p)\sinh(\pi b/p)}\bigg)^{-\theta/2\pi i}\\
\times &\bigg[\sinh\bigg(\frac{\theta}{p}\bigg)\bigg(\frac{\sinh(\pi a/p)\sinh(\pi c/p)}{\sinh (\pi d/p)\sinh(\pi b/p)}\bigg)^{1/2}\delta_{db}+\sinh\bigg(\frac{i\pi-\theta}{p}\bigg)\delta_{ac}\bigg],
\end{align*}
where $a=\overline{1,k+1}$ is the vacuum index, $p=k+2$ and
\begin{align*}
u(\theta)&=\Gamma\bigg(\frac{1}{p}\bigg)\Gamma\bigg(1+\frac{i\theta}{p}\bigg)\Gamma\bigg(1-\frac{\pi+i\theta}{p}\bigg)\prod_{n=1}^\infty\frac{R_n(\theta)R_n(i\pi-\theta)}{R_n(0)R_n(i\pi)},\\
R_n(\theta)&=\frac{\Gamma(2n/p+i\theta/\pi p)\Gamma(1+2n/p+i\theta/\pi p)}{\Gamma((2n+1)/p+i\theta/\pi p)\Gamma(1+(2n-1)/p+i\theta/\pi p)}\;.
\end{align*}
The scattering matrix is the tensor product of the $SU(2)$ chiral Gross-Neveu S-matrix and  the kink S-matrix 
\begin{align}
\mathcal{S}^{\text{SU(2)}_k}(\theta)= \mathcal{S}_{[k]}^\text{kink}(\theta)\otimes\mathcal{S}^{\text{SU(2)}}(\theta).\label{level-k-s-matrix}
\end{align}
The TBA system for perturbed $SU(2)_k$ consists of two parts. The right wing consists of $SU(2)$ magnon bound states, exactly like the Gross-Neveu model. The left wing are formed of kink magnon bound states. There are a priori $k$ of them but the longest one does not contribute to the thermodynamic properties. This results in a reduced TBA system involving only $k-1$ kink magnons
\begin{align}
\log Y_n(u)+Rm\cosh(u)\delta_{n,k}=\sum_{m=0}^\infty K_{mn}\star\log(1+Y_m)(u),\quad n=\overline{1,\infty}.\label{level-k-TBA}
\end{align}
\begin{figure}[h]
\centering
\includegraphics[width=14cm]{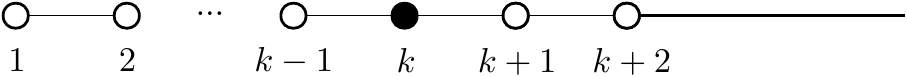}
\caption{Y system for current perturbed $SU(2)_k$ WZNW. The $k$'th node is the physical rapidity. The $j$'th node is kink magnon string of length $k-j$ for $1\leq j\leq k-1$. The $a$'th node is $SU(2)$ magnon string of length $a-k$ for $a\geq k+1$.}
\end{figure}

The scattering kernels are given at the end of appendix \ref{higher-reduced}. For later discussion of  g-function we present here their convolutions with identity
\begin{gather*}
K_{ij}\star\mathbf{1}=\delta_{ij}-2\frac{\min(i,j)[k-\max(i,j)]}{k},\quad K_{ab}\star \mathbf{1}=\delta_{ab}-2\min(a-k,b-k),\\
K_{kj}\star\mathbf{1}=-K_{jk}\star\mathbf{1}=-\frac{j}{k},\quad K_{ka}\star\mathbf{1}=-K_{ak}\star\mathbf{1}=-1,\quad i,j\in\overline{1,k-1},a,b\in\overline{k+1,\infty},\\
K_{kk}\star\mathbf{1}=1-\frac{1}{2k}.
\end{gather*}
The TBA equations \eqref{level-k-TBA} can be transformed into an equivalent Y system.  The same kernel $s$ in \eqref{Fourier-s} connects the two wings to the physical node, despite different scattering structures on each wing. Again, we denote $\mathcal{Y}_n=Y_n^{2\delta_{n,k}-1}$
\begin{align}
\log \mathcal{Y}_n+RE\delta_{n,k}= s\star[\log(1+\mathcal{Y}_{n-1})+\log(1+\mathcal{Y}_{n+1})],\quad n=\overline{1,\infty}\label{level-k-Y-system}
\end{align}
The UV and IR solutions of this Y-system are
\begin{gather*}
1+\mathcal{Y}_n^{\text{UV}}=(n+1)^2,\quad n=\overline{1,\infty},\\
1+\mathcal{Y}_n^{\text{IR}}=\frac{\sin^2[(n+1)\pi/(k+2)]}{\sin^2[\pi/(k+2)]},\quad n=\overline{1,k-1},\quad 1+\mathcal{Y}_n^{\text{IR}}=(n-k+1)^2,\quad n=\overline{k,\infty},
\end{gather*}
from which we recover the central charge of the unperturbed CFT 
\begin{align*}
\sum_{n=1}^\infty\textnormal{Li}_R\big(\frac{1}{1+\mathcal{Y}_n^{\text{IR}}}\big)-\textnormal{Li}_R\big(\frac{1}{1+\mathcal{Y}_n^{\text{UV}}}\big)=\frac{3k}{k+2}.
\end{align*}
For a proof of this identity, see for instance \cite{Kirillov:1993ih}. We can add to the TBA equations \eqref{level-k-TBA} a chemical potential coupled to the $SU(2)$ symmetry only. The Y system \eqref{level-k-Y-system} is again protected. 
In the IR limit the left wing decouples from the right wing and is immune to the $SU(2)$ chemical potential. The IR values on the left wing is therefore unchanged, while on the right wing we have
\begin{align*}
1+\mathcal{Y}_n^{\text{IR}}(\k)=[n-k+1]_\k^2,\quad n=\overline{k,\infty}.
\end{align*}
In the UV limit all nodes are affected by the twist
\begin{align*}
1+\mathcal{Y}_n^{\text{UV}}(\k)=[n+1]_\k,\quad n=\overline{1,\infty}.
\end{align*}
Similar to the case of level 1, we can compute the particle densities in this limit
\begin{gather}
\lim_{R\to 0} \pi RD_0(R,\mu)=2\log \frac{1-\k^{k+1}}{1-\k}-k\log \k,\quad \lim_{R\to 0}\sum_{a=1}^\infty \pi R a D_a(R,\mu)=2\log \frac{1-\k^{k+1}}{1-\k}
 \label{particle-densities-level-k}
\end{gather}
as well as the scaled free energy density
\begin{align}
\lim_{R\to 0}c(R,\mu)=\frac{3k}{k+2}-\frac{6k\mu^2}{\pi^2}.
\end{align}
\section{The reflection factors}
Due to the factorization property of the S-matrix \eqref{level-k-s-matrix}, we can study the reflection factors for kink magnons and $SU(2)$ magnons independently. 

After the maximum string reduction procedure (see appendix \eqref{higher-reduced}), the effective scattering between $k-1$ kink magnon strings are very similar to the scattering of the A theories in the ADE family \cite{Klassen:1989ui}. The scattering phase between a kink magnon string of length $n$ and another one of length $m$, for $n,m=\overline{1,k-1}$ is given by
\begin{align}
S_{nm}(\theta)=\frac{\sinh\big(\frac{\theta}{k}+i\pi\frac{|n-m|}{2k}\big)}{\sinh\big(\frac{\theta}{k}-i\pi\frac{|n-m|}{2k}\big)}\frac{\sinh\big(\frac{\theta}{k}+i\pi\frac{n+m}{2k}\big)}{\sinh\big(\frac{\theta}{k}-i\pi\frac{n+m}{2k}\big)}\prod_{s=\frac{|n-m|}{2}+1}^{\frac{n+m}{2}-1}\bigg[\frac{\sinh\big( \frac{\theta}{k}+i\pi \frac{s}{k}\big)}{\sinh\big( \frac{\theta}{k}-i\pi \frac{s}{k}\big)}\bigg]^2\label{kink-magnon-scatterings}
\end{align}
where $\theta$ is the rapidity difference between the two string centers.
On the other hand, the $A_{k-1}$ S-matrix describes the purely elastic scattering of the coset CFT $SU(k)_1\times SU(k)_1/SU(k)_2$ ($\mathbb{Z}_{k}$ parafermions) perturbed by its $(1,1,\text{adj})$ operator \footnote{The central charge of this CFT is exactly the central charge of $SU(2)_k$ minus 1  : parafermion $\mathbb{Z}_k$ can be represented as $SU(2)_k/U(1)$.}. This massive perturbation consists of $k-1$ particles $n=1,...,k-1$ where $\bar{n}=k-n$ with mass spectrum
\begin{align*}
m_n=m\sin\big(\frac{\pi n}{k}\big)/\sin\big(\frac{\pi}{k}\big).
\end{align*}
The purely elastic scattering between these particles is
\begin{gather}
S_{nm}(\theta)=\frac{\sinh\big(\frac{\theta}{2}+i\pi\frac{|n-m|}{2k}\big)}{\sinh\big(\frac{\theta}{2}-i\pi\frac{|n-m|}{2k}\big)}\frac{\sinh\big(\frac{\theta}{2}+i\pi\frac{n+m}{2k}\big)}{\sinh\big(\frac{\theta}{2}-i\pi\frac{n+m}{2k}\big)}\prod_{s=\frac{|n-m|}{2}+1}^{\frac{n+m}{2}-1}\bigg[\frac{\sinh\big( \frac{\theta}{2}+i\pi \frac{s}{k}\big)}{\sinh\big( \frac{\theta}{2}-i\pi \frac{s}{k}\big)}\bigg]^2\label{A-series-scatterings}
\end{gather}
So indeed, despite the different underlying physics, the two scattering phases \eqref{kink-magnon-scatterings} and \eqref{A-series-scatterings} are identical up to a redefinition of rapidity variable
\begin{align*}
\theta^{\text{ kink magnon}}=\frac{k}{2}\times\theta^{\text{ A series}}.
\end{align*}
This suggests that we can use the minimal reflection factor already derived for A series \cite{Corrigan:1994ft},\cite{Dorey:2005ak} for the kink magnons. It is the solution of the boundary unitarity, crossing and bootstrap equations with a minimal number of poles and zeros
\begin{align*}
R_j(\theta)=\prod_{s=0}^{j-1}\frac{\sinh\big(\frac{\theta}{2}+i\pi\frac{s}{2}\big)}{\sinh\big(\frac{\theta}{2}-i\pi\frac{s}{2}\big)}\frac{\sinh\big(\frac{\theta}{2}-i\pi\frac{k-s+1}{2}\big)}{\sinh\big(\frac{\theta}{2}+i\pi\frac{k-s+1}{2}\big)},\quad j=\overline{1,k-1}.
\end{align*}
It satisfies in particular the following identity
\begin{align}
K_j\star \mathbf{1}-\frac{1}{2}K_{jj}\star\mathbf{1}-\frac{1}{2}=0,\quad j=\overline{1,k-1}.\label{A-series-boundary}
\end{align}
where $K_j=-i\partial\log R_j$, which greatly simplifies the form of the corresponding g-function in the next section. For parafermions, this set of relection factors is assigned with the fixed boundary condition or equivalently the vacuum representation of both $SU(k)_1$ and $SU(k)_2$. The g-function was found to be
\begin{align}
g_0^2=\frac{2}{\sqrt{k+2}\sqrt{k}}\sin\frac{\pi}{k+2}.\label{A-series-g-function}
\end{align}
On the other hand, we consider trivial reflection factors on the $SU(2)$ magnons. We do not aim at proving this point but we merely conjecture it based on the result of non linear $O(N)$ sigma model with boundary \cite{Aniceto:2017jor}, \cite{Gombor:2017qsy}, where similar magnon structure arises. What we are doing is to first diagonalize the bulk theory by nested Bethe Ansatz technique. We then treat the theory as one with diagonal scattering and find the reflection factors based on this bulk diagonal scattering. The standard way to do it would be to start with a set of reflection factors that satisfy the boundary Yang-Baxter equation. One then writes Bethe equations with these reflection factors and diagonalizes the corresponding two-row transfer matrix. 

To summarize, we conjecture the following set of TBA equations for current perturbed $SU(2)_k$ theories in the presence of boundaries
\begin{gather*}
\text{physical rapidity}\quad e^{iLm\sinh(\theta_{k,n})}R_{k}^2(\theta_{k,n})\prod_{j=1}^\infty\prod_{m}S_{kj}(\theta_{k,n}-\theta_{j,m})S_{kj}(\theta_{k,n}+\theta_{j,m})=-1\\
\text{Kink magnons}\quad R_{j}^2(\theta_{j,n})\prod_{l=1}^k\prod_{m}S_{jl}(\theta_{j,n}-\theta_{l,m})S_{jl}(\theta_{j,n}+\theta_{l,m})=-1,\quad j=\overline{1,k-1}\\
\text{SU(2) magnons}\quad \prod_{l=k}^\infty\prod_{m}S_{jl}(\theta_{j,n}-\theta_{l,m})S_{jl}(\theta_{j,n}+\theta_{l,m})=-1,\quad j=\overline{k+1,\infty}
\end{gather*}
We denote from now on the convolution of the boundary reflections with identity by $B_j=K_j\star \mathbf{1}$. They are given by \eqref{A-series-boundary} for kink magnons and are zero for $SU(2)$ magnons. For the physical rapidity, we leave it as a parameter.
\section{g function}
We now have all the necessary ingredients to study the UV and IR limit of the g-function of the current-perturbed $SU(2)_k$ theories.

For convenience we repeat here the result (1)-(3), with an equivalent form of the loop part that is more adapted to actual computation
\begin{align}
&2\log g(R)=2\log g^\textnormal{trees}(R)+2\log g^\textnormal{loops}(R),\label{g-function}\\
&2\log g^\textnormal{trees}(R)=\sum_{n}\int_{-\infty}^\infty\frac{du}{2\pi}[K_{n}(u)-K_{nn}(u,-u)-\pi\delta(u)]\log[1+Y_n(u)],\label{tree-part}\\&2\log g^\textnormal{loops}(R)\nonumber\\
&=\sum_{n\geq 1}\frac{1}{n}\sum_{a_1,...,a_n\geq 0}\bigg[\prod_{j=1}^n\int\limits_{-\infty}^{+\infty}\frac{du_j}{2\pi}f_{a_j}(u_j)\bigg]K_{a_1a_2}(u_1+u_2)K_{a_2a_3}(u_2-u_3)...K_{a_na_1}(u_n-u_1).\label{equivalent-expression}
\end{align}
The goal of this section is to support to our proposition \eqref{proposition} by proving that it is possible to match the normalized UV g-function, namely $g_{\text{UV}}/g_{\text{IR}}$ with a conformal g-function \eqref{CFT-entropy} in some cases. While carrying out this normalization we encounter divergence in both IR and UV limit. We illustrate this phenomenon for the Gross-Neveu model and show how an appropriate regularization could lead to a finite ratio.
\subsection{Level 1- Gross Neveu model}
At zero temperature, the tree part \eqref{tree-part} of the g-function can be exactly evaluated. With the Y-functions given by constants in \eqref{IR-GN-not-twist}, the reflection kernels for $SU(2)$ magnons being zero and the scattering kernels $K_{nn}$ given in \eqref{aux-aux}, it turns out to be divergent in this limit
\begin{align}
2\log g^\textnormal{trees}_{\text{IR}}=\sum_{n=1}^\infty(n-1)\log\big[1+\frac{1}{n(n+2)}\big].\label{tree-GN-not-twist}
\end{align}
The tadpole (the $n=1$ term in the series \eqref{equivalent-expression}) suffers from a similar divergence 
\begin{align}
2\log g^\textnormal{tadpole}_{\text{IR}}&=\sum_{n=1}^\infty\frac{Y_n^{\text{IR}}}{1+Y_n^{\text{IR}}}\int_{-\infty}^{+\infty}\frac{du}{2\pi}K_{nn}(2u)=\sum_{n=1}^\infty\frac{1-2n}{2(n+1)^2}.\label{tadpole-GN}
\end{align}
This logarithmic divergence is present for higher order terms and for the infinite temperature limit alike. We believe it is a common feature among models with an infinite number of string magnons. 

As a regularization, we propose to use the twisted TBA solutions \eqref{IR-UV-GN-twist}. The tree part of the IR g-function can now be expressed in terms of the twist parameter $\k$
\begin{align}
2\log g^\textnormal{trees}_{\text{IR}}(\k)
=\sum_{n=1}^\infty(n-1)\log\big(1+\frac{1}{[n+1]_{\k}^2-1}\big)=-\log(1-\k^2).\label{g-GN-IR-tree}
\end{align}
To evaluate the loop part, we remark that for constant Y-functions the series \eqref{equivalent-expression} can be written as a determinant 
\begin{align}
2\log g^{\textnormal{loops}}_{\text{IR}}(\k)=-\frac{1}{2}\log\det[1-\hat{K}^{\text{IR}}(\k)],
\end{align}
where 
\begin{gather}
\hat{K}_{ab}^{\text{IR}}(\k)\equiv \sqrt{\frac{Y_a^{\text{IR}}(\k)}{1+Y_a^{\text{IR}}(\k)}\frac{Y_b^{\text{IR}}(\k)}{1+Y_b^{\text{IR}}(\k)}}\int_{-\infty}^{+\infty}\frac{du}{2\pi}K_{ab}(u),\label{first-Fredholm}
\end{gather}
The factor $1/2$ comes from the change of variables $(u_1+u_2,u_2-u_3,...,u_n-u_1)\to (\tilde{u}_1,\tilde{u}_2,...,\tilde{u}_n)$. We show in appendix \ref{det-IR} that 
\begin{align}
\det[1-\hat{K}^\text{IR}(\k)]=(1-\k)^{-1}.\label{g-GN-IR-loop}
\end{align}
By combining the two contributions \eqref{g-GN-IR-tree} and \eqref{g-GN-IR-loop}, we obtain the IR g-function of Gross-Neveu model as a function of the twist parameter. In the untwisted limit $\k\to 1$ it behaves as
\begin{align}
\lim_{\k\to 1}2\log g_{\text{IR}}(\k)=-\log 2-\frac{1}{2}\log(1-\k).\label{g-GN-IR}
\end{align}

We can repeat the same analysis for the UV limit, using the corresponding twisted constant solution \eqref{IR-UV-GN-twist}. Compared to the IR limit we algo get contribution from the physical rapidity. The loop part can again be written as a determinant by replacing the IR by UV values in the matrix \eqref{first-Fredholm}. We show in appendix \ref{det-UV} that this determinant is again a very simple function of the twist parameter
\begin{align}
2\log g^\textnormal{trees}_{\text{UV}}(\k)&=(B_0-\frac{3}{4})\log\frac{(1+\k)^2}{\k}-\log(1-\k^3),\\
2\log g^\textnormal{loops}_{\text{UV}}(\k)&=\frac{1}{2}\log[2(1-\k)].
\end{align}
The UV value of g-function exhibits the same divergence as the IR one in the untwisted limit 
\begin{align}
\lim_{\k\to 1}2\log g_{\text{UV}}(\k)=(2B_0-1)\log 2-\log 3-\frac{1}{2}\log(1-\k)\label{g-GN-UV}
\end{align}
In particular their ratio is well defined
\begin{align}
\bigg(\frac{ g_{\text{UV}}}{g_{\text{IR}}}\bigg)^2=2^{2B_0}/3.\label{final-level-1}
\end{align}
The two Cardy g-functions \eqref{CFT-entropy} of $SU(2)_1$ CFT take the same value $g_1^2=g_2^2=1/\sqrt{2}$. For integrable boundary conditions, the reflection factor usually gives rational value for $B_0$ and our proposition \eqref{final-level-1} could not be matched with a Cardy g-function. We carry on our analysis to higher levels.
\subsection{Higher levels}
We first consider the IR limit, in which the left and right wing are decoupled. The former is not affected by the twist while the latter is identical to the IR of the Gross-Neveu model. Our choice of reflection factors with the property \eqref{A-series-boundary} eliminates the left wing from the tree part of the g-function. As a consequence we get the same result as the IR tree part of  Gross-Neveu model \eqref{g-GN-IR-tree}
\begin{align}
2\log g_{\text{IR}}^\textnormal{trees}(\k)=-\log(1-\k^2).\label{tree-IR-level-n}
\end{align}
The loop part is factorized into two determinants
\begin{align}
2\log g_{\text{IR}}^\textnormal{loops}=-\frac{1}{2}\log\det(1-\hat{K}_{1\to k-1})-\frac{1}{2}\log\det(1-\hat{K}_{k+1\to \infty}).
\end{align}
The finite  determinant involving the trigonometric Y-functions has been computed in \cite{Dorey:2005ak} while the infinite  determinant is exactly the same as that of IR Gross-Neveu 
\begin{align*}
\det(1-\hat{K}_{1\to k-1})=[\frac{4k}{k+2}\sin^2\frac{\pi}{k+2}]^{-1}
,\quad \det(1-\hat{K}_{k+1\to \infty})=(1-\kappa)^{-1}.
\end{align*}
By summing the two parts, we obtain the IR  g-function. Its behavior in the untwisted limit is
\begin{align}
\lim_{\k\to 1}2\log g_{\text{IR}}(\k)=-\log 2+\frac{1}{2}\log\frac{4k}{k+2}+\log\sin\frac{\pi}{k+2}-\frac{1}{2}\log(1-\k).\label{IR-level-k}
\end{align}

In the UV limit all Y-functions are twisted. Again only the right wing contributes to the tree part of g-function
\begin{align}
2\log g_{\text{UV}}^\textnormal{trees}(\k)=(B_k-1+\frac{1}{4k})\log [k+1]_\k^2-\log(1-\k^{k+2}).\label{tree-g-level-n}
\end{align}
The loop contribution is given by a determinant that connects the two wings. We compute this  determinant in appendix \ref{det-UV}.  Despite its complicated form, as the structure of scattering kernels on the left and right wing are different, the result is simple
\begin{align}
2\log g^\textnormal{loops}_{\text{UV}}(\k)=\frac{1}{2}\log 2k+\frac{1}{2}\log (1-\k).\label{loop-g-level-n}
\end{align}
The UV g-function is obtained by summing \eqref{tree-g-level-n} and \eqref{loop-g-level-n}
\begin{align}
\lim_{\k\to 1}2\log g_{\text{UV}}(\k)=(2B_k-2+\frac{1}{2k})\log (k+1)-\log(k+2)+\frac{1}{2}\log 2k-\frac{1}{2}\log (1-\k).\label{UV-level-k}
\end{align}

We see that the IR \eqref{IR-level-k} and UV \eqref{UV-level-k} values of g-function exhibit the same divergence in the untwisted limit. We can therefore extract their ratio
\begin{align}
\bigg(\frac{g_{\text{UV}}}{g_{\text{IR}}}\bigg)^2=(k+1)^{2B_k-2+\frac{1}{2k}}\times\sqrt{\frac{2}{k+2}}\times\frac{1}{\sin\frac{\pi}{k+2}}.\label{a-song-of-ice-and-fire}
\end{align}
To remind, the Cardy g-functions are given by
\begin{align}
g_\lambda^2=\sqrt{\frac{2}{k+2}}\times\frac{1}{\sin\frac{\pi}{k+2}}\times\sin^2\frac{(\lambda+1)\pi}{k+2},\;0\leq\lambda\leq k\label{g-cardy-level-n}
\end{align}
Therefore we can match our normalized UV g-function \eqref{a-song-of-ice-and-fire} with $g_{k/2}$ for even $k$ as long as the reflection factor of the physical rapidity satisfies $B_k=1-1/(4k)$. Let $k=2m$ then the corresponding bulk primary has conformal dimension
\begin{align*}
\Delta=\frac{m(m+2)}{8(m+1)}.
\end{align*}
\section{Conclusion}
In this paper we propose the following procedure to study the g-function of a massive integrable theory with non-diagonal bulk scattering
\begin{itemize}
\item Diagonalize the theory using the Nested Bethe Ansatz technique.
\item Treat the newly obtained theory as diagonal with extra magnonic particles and apply the results \eqref{trees+loops}-\eqref{loops} to compute its g-function.
\item Normalize the g-function by its zero temperature limit value.
\end{itemize}
We test our proposition for the current-perturbed $SU(2)$ WZNW CFTs. The TBA of these theories involves an infinite tower of magnon strings. As a consequence both the tree \eqref{trees} and loop \eqref{loops} part of the g-function diverge at zero and infinite temperature. This phenomenon is illustrated for the Gross-Neveu model in \eqref{tree-GN-not-twist},\eqref{tadpole-GN}. We conjecture that such divergence is present at arbitrary temperature. By considering the twisted TBA, we are able to compute these two limits of g-function as functions of the twist parameter $\k$. It is found that they exhibit the same divergence $-\frac{1}{2}\log(1-\k)$  in the untwisted  limit $\k\to 1$ \eqref{IR-level-k},\eqref{UV-level-k}. The normalized UV g-function is then well defined \eqref{a-song-of-ice-and-fire} and can be identified with a Cardy g-function of the unperturbed CFT under some assumption on the reflection factor of the physical rapidity and for even levels.

This normalization has a diagramatical interpretation in the formulation of \cite{Kostov:2018dmi}. At zero temperature the boundary entropy is given by the sum of all graphs made exclusively of auxiliary magnons. The contribution of these graphs does not depend on the temperature and can be absorbed into the normalization of the partition function. No physical observable will involve such graphs.
\section*{Acknowledgement}
The authors thank Benjamin Basso, Romuald Janik, Hubert Saleur, Patrick Dorey, Paul Fendley, Zoltan Bajnok, Tamas Gombor, Changrim Ahn, Stijn van Tongeren for valuable help, discussions and suggestions during various stages of the draft.
\appendix
\section{Gross-Neveu TBA and Y system}
\label{GN-TBA}
The Bethe equations for a state of $N$ physical rapidities $\theta_1,...,\theta_N$ and $M$ magnonic rapidities $u_1,...,u_M$ read
\begin{align*}
1&=e^{ip(\theta_j)L}\prod_{k\neq j}^NS_0(\theta_j-\theta_k)\prod_{m=1}^M\frac{\theta_j-u_m+i\pi/2}{\theta_j-u_m-i\pi/2}\\
1&=\prod_{j=1}^N\frac{u_k-\theta_j-i\pi/2}{u_k-\theta_j+i\pi/2}\prod_{l\neq k}^M\frac{u_k-u_l+i\pi}{u_k-u_l-i\pi}
\end{align*}
String solutions are formed of magnon rapidities equally spaced in distance of $i$. Let $u_{k,n}$ be the real center of a string of length $n$ then the ensemble of string rapidities are given by
\begin{align*}
u_{k,n}^a=u_{k,n}-i\pi\frac{n+1}{2}+i\pi a, \quad a=1,...,n
\end{align*}
The scattering phase between strings (and physical node) is the product between the scattering phases of their constituents
\begin{gather*}
S_{0n}(\theta,u_{k,n})=\frac{\theta-u_{k,n}+i\pi n/2}{\theta-u_{k,n}-i\pi n/2},\\
S_{nm}(u_{k,n},u_{l,m})=\frac{u_{k,n}-u_{l,m}+i\pi\frac{|n-m|}{2}}{u_{k,n}-u_{l,m}-i\pi\frac{|n-m|}{2}}\frac{u_{k,n}-u_{l,m}+i\pi\frac{n+m}{2}}{u_{k,n}-u_{l,m}-i\pi\frac{n+m}{2}}\prod_{s=\frac{|n-m|}{2}+1}^{\frac{n+m}{2}-1}\bigg[\frac{u_{k,n}-u_{l,m}+i\pi s}{u_{k,n}-u_{l,m}-i\pi s}\bigg]^2.
\end{gather*}
The corresponding scattering kernels are
\begin{gather}
K_{0n}(\theta-u)=-K_{n,0}(\theta-u)=-\frac{4\pi n}{4(\theta-u)^2+\pi^2n^2},\label{kernel-0-n}\\
K_{nm}(u)=K_{m,n}(u)=(1-\delta_{nm})K_{0,|n-m|}(u)+K_{0,n+m}(u)+2\sum_{s=\frac{|n-m|}{2}+1}^{\frac{n+m}{2}-1}K_{0,2s}(u).\label{kernel-n-m}
\end{gather}
Their Fourier transforms are simple
\begin{gather}
\hat{K}_{0n}(w)=-\hat{K}_{n0}(w)=-e^{-\pi n|w|/2}\label{phy-aux}\\
\hat{K}_{nm}(w)=\delta_{nm}+(e^{\pi|w|}+1)\frac{e^{-(n+m)\pi|w|/2}-e^{-|n-m|\pi|w|/2}}{e^{\pi|w|}-1}\label{aux-aux}
\end{gather}
Here we normalize the Fourier transformation as
\begin{align*}
\hat{f}(w)=\frac{1}{2\pi}\int_{-\infty}^{+\infty}f(t)e^{iwt}dt.
\end{align*}
For the physical-physical scattering \eqref{scalar-part}
\begin{gather}
K_{00}(\theta)=\frac{1}{\pi}\sum_{l=0}^\infty-\frac{l+1}{(l+1)^2+\theta^2/4\pi^2}+\frac{l+1/2}{(l+1/2)+\theta^2/4\pi^2}\Rightarrow\hat{K}_{00}(w)=\frac{e^{-\pi|w|/2}}{2\cosh(\pi w/2)}\label{phy-phy}
\end{gather}
The above kernels control the TBA equations
\begin{align}
Y_n(u)=e^{-\delta_{0n}RE(\theta)}\exp\big[\sum_{m\geq 0}K_{m,n}\star\log(1+Y_m)(u)\big].\label{TBA}
\end{align}
By defining $\mathcal{Y}_n=Y_n^{-1}$ for $n\geq 1$ and $\mathcal{Y}_0=Y_0$ we can transform this to the Y-system  
\begin{align}
\log \mathcal{Y}_n+\delta_{n0}RE=\sum_{m=0}^\infty I_{mn}s\star\log(1+\mathcal{Y}_m).\label{Y}
\end{align}
where the kernel $s$ has the following Fourier transform
\begin{align}
\hat{s}(w)=\frac{1}{2\cosh(\pi w/2)}.\label{Fourier-s}
\end{align}
To prove that \eqref{TBA} leads to \eqref{Y} we first act by $-s$ to the TBA equation of $Y_1$
\begin{align}
\log Y_1&=K_{01}\star\log(1+Y_0)+K_{11}\star\log(1+Y_1)+\sum_{n\geq 2}K_{n1}\star\log(1+Y_n).\label{Y1}
\end{align}
With help of the following identities
\begin{gather*}
-s\star K_{01}=K_{00},\quad -s\star K_{11}=K_{1,0}-s,\quad s\star K_{n1}=K_{0n},\quad n\geq 2.
\end{gather*}
We can write \eqref{Y1} as
\begin{align*}
\log Y_0+RE=s\star\log (1+\frac{1}{Y_1})
\end{align*}
which is the first equation of Y system. Next, we act $s$ to the TBA equation of $Y_2$
\begin{align*}
\log Y_2&=K_{02}\star\log(1+Y_0)+K_{12}\star\log(1+Y_1)+K_{22}\star\log(1+Y_2)+\sum_{n\geq 3}K_{n2}\star\log(1+Y_n).
\end{align*}
this time we need the folowing identities
\begin{gather*}
s\star K_{02}=K_{01}+s,\quad s\star K_{12}=K_{11},\quad s\star K_{22}=K_{21}+s,\quad s\star K_{n2}=K_{n1},\quad n\geq 3.
\end{gather*}
From which we have
\begin{align*}
\log \mathcal{Y}_1 =s\star\log(1+Y_0)+s\star\log(1+\mathcal{Y}_2).
\end{align*}
For $n\geq 2$ we can show from the average property $s\star(K_{0,n-1}+K_{0,n+1})=K_{0n}$ that
\begin{align*}
s\star\log Y_{n+1}+s\star \log Y_{n-1}=\log Y_n+s\star\log(1+Y_{n+1})+s\star\log(1+Y_{n-1}).
\end{align*}
\section{Derivation for higher levels}
\label{higher-TBA}
\subsection{The scattering and the kernels}
The scalar factors and their exponential form \cite{HOLLOWOOD199443}
\begin{gather*}
S_0^{SU(2)}(\theta)=-\frac{\Gamma(1-\theta/2\pi i)}{\Gamma(1+\theta/2\pi i)}\frac{\Gamma(1/2+\theta/2\pi i)}{\Gamma(1/2-\theta/2\pi i)},\\
S_0^{[k]}(\theta)=\exp\bigg[\int_{-\infty}^{+\infty}\frac{dx}{x}e^{2i\theta x/\pi}\frac{\sinh[(k+1)x]\sinh x}{\sinh [(k+2)x]\sinh (2x)}\bigg].
\end{gather*}
The Bethe equations involving $N$ physical rapidities $\theta$, $M$ $SU(2)$ magnon rapidities $u$ and $P$ kink magnon rapidities $v$
\begin{gather*}
e^{-ip(\theta_j)L}=-\epsilon_j\prod_{i=1}^N S_0^{SU(2)}(\theta_j,\theta_i)S_0^{[k]}(\theta_j,\theta_i)\prod_{k=1}^M\frac{\theta_j-u_k+i\pi/2}{\theta_j-u_k-i\pi/2}\prod_{q=1}^P\frac{\sinh \dfrac{\theta_j-v_q+i\pi/2}{k+2}}{\sinh\dfrac{\theta_j- v_q-i\pi/2}{k+2}},\\
\prod_{j=1}^N\frac{u_k-\theta_j+i\pi/2}{u_k-\theta_j-i\pi/2}=\Omega_k\prod_{l=1}^M\frac{u_k-u_l+i\pi}{u_k-u_l-i\pi},\\
\prod_{j=1}^N\frac{\sinh\dfrac{v_q-\theta_j+i\pi/2}{k+2}}{\sinh\dfrac{v_q-\theta_j-i\pi/2}{k+2}}=\Omega_q\prod_{p=1}^P\frac{\sinh\dfrac{v_q-v_p+i\pi}{k+2}}{\sinh\dfrac{v_q-v_p-i\pi}{k+2}}.
\end{gather*}
with some constants $\epsilon_j,\Omega_k,\Omega_q$.
String solutions
\begin{gather*}
\text{ u strings of length } n=\overline{1,\infty}: \quad u_{k,n}^a=u_{k,n}-i\pi\frac{n+1}{2}+i\pi a, \quad a=1,...,n\\
\text{ v strings of length } m=\overline{1,k}: \quad v_{q,m}^b=v_{q,m}-i\pi\frac{m+1}{2}+i\pi b, \quad b=1,...,m
\end{gather*}
The scatterings between $SU(2)$ strings with themselves and between them and the physical rapidity are the same as before.
For kink magnon strings
\begin{gather*}
S_{0n}^{[k]}(\theta,v_{q,n})=\frac{\lbrace\theta-v_{q,m}+i\pi m/2\rbrace_k}{\lbrace\theta-v_{q,m}-i\pi m/2\rbrace_k}\\
S_{nm}^{[k]}(v_{q,n},v_{p,m})=\frac{\lbrace v_{q,n}-v_{p,m}+i\pi\frac{|n-m|}{2}\rbrace_k}{\lbrace v_{q,n}-v_{p,m}-i\pi\frac{|n-m|}{2}\rbrace_k}\frac{\lbrace v_{q,n}-v_{p,m}+i\pi\frac{n+m}{2}\rbrace_k}{\lbrace v_{q,n}-v_{p,m}-i\pi\frac{n+m}{2}\rbrace_k}\prod_{s=\frac{|n-m|}{2}+1}^{\frac{n+m}{2}-1}\bigg[\frac{\lbrace v_{q,n}-v_{p,m}+i\pi s\rbrace_{k}}{\lbrace v_{q,n}-v_{p,m}-i\pi s\rbrace_{k}}\bigg]^2
\end{gather*}
where we have noted for convenience
\begin{align*}
\lbrace u\rbrace_k  =\sinh\frac{u}{k+2}.
\end{align*}
The Fourier transforms of the kink magnon strings scattering kernel
\begin{gather*}
\hat{K}_{0n}^{[k]}(w)=-\frac{\sinh[(k+2-n)\frac{\pi w}{2}]}{\sinh(k+2)\frac{\pi w}{2}},\\
\hat{K}_{nm}^{[k]}(w)=\delta_{nm}-2\frac{\sinh\big[ \min(n,m)\frac{\pi w}{2}\big]\sinh\big[ (k+2-\max(n,m))\frac{\pi w}{2}\big]\cosh\frac{\pi w}{2}}{\sinh\big[(k+2)\frac{\pi w}{2}\big]\sinh\frac{\pi w}{2}}.
\end{gather*}
\subsection{Maximal string reduction and reduced TBA}
\label{higher-reduced}
At this point we have the raw TBA equations
\begin{gather*}
\log Y_{\tilde{n}}=\sum_{m=0}^k K_{mn}^{[k]}\star\log(1+Y_{\tilde{m}}),\quad n=\overline{1,k},\\
\log Y_0+RE=\sum_{n=0}^kK_{n0}^{[k]}\log(1+Y_{\tilde{n}})+\sum_{n=1}^\infty K_{n0}^{\text{SU(2)}}\log(1+Y_n),\\
\log(1+Y_n)=\sum_{m=0}^\infty K_{mn}\star\log(1+Y_m),\quad n=\overline{1,\infty}.
\end{gather*}
where we have used the tilde indices to denote kink rapidities, also $\tilde{0}=0$.
\begin{figure}[h]
\centering
\includegraphics[width=14cm]{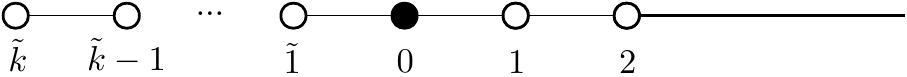}
\end{figure}

Maximal string reduction \cite{HOLLOWOOD199443}: $u$ string of length $k$ doesn't contribute in the thermodynamic limit. The $\tilde{k}$ string is frozen in the sense that $Y_{\tilde{k}}=\infty$. We look at the TBA equation for this string
\begin{align*}
\log Y_{\tilde{k}}=\sum_{m=0}^kK_{mk}^{[k]}\star\log(1+Y_{\tilde{m}}).
\end{align*}
Upon replacing $\log Y_{\tilde{k}}$ by $\log(1+Y_{\tilde{k}})$, we can effectively remove the $\tilde{k}$ node from our TBA system
\begin{align*}
\log(1+Y_{\tilde{k}})=\sum_{m=0}^{k-1}(1-K_{kk}^{[k]})^{-1}\star K_{mk}^{[k]}\star\log(1+Y_{\tilde{m}}).
\end{align*}
The reduced system (only the kink magnon related part) read
\begin{align*}
\log Y_{\tilde{n}}&=\sum_{m=0}^{k-1}\bigg[K_{kn}^{[k]}\star(1-K_{kk}^{[k]})^{-1}\star K_{mk}^{[k]}+K_{mn}^{[k]}\bigg]\star\log(1+Y_{\tilde{m}}),\quad n=\overline{1,k-1}\\
\log Y_0+RE &=\sum_{n=0}^{k-1}\bigg[K_{k0}^{[k]}\star(1-K_{kk}^{[k]})^{-1}\star K_{nk}^{[k]}+K_{n,0}^{[k]}\bigg]\star\log(1+Y_{\tilde{n}})+.....
\end{align*}
The following identity drastically simplifies this system
\begin{align}
K_{kn}^{[k]}\star(1-K_{kk}^{[k]})^{-1}\star K_{mk}^{[k]}+K_{mn}^{[k]}=K_{mn}^{[k-2]},\quad m,n=\overline{0,k-1}.\label{reduction-identity}
\end{align}
To summarize, the reduced TBA system for integrable perturbed $SU(2)_k$ is 
\begin{align*}
\log Y_n+\delta_{n,k}RE=\sum_{m,n} K_{mn}\star\log(1+Y_m),\quad n=\overline{1,\infty}
\end{align*}
where
\begin{gather*}
\hat{K}_{kn}(w)=-\hat{K}_{nk}(w)=-\frac{\sinh n\frac{\pi w}{2}}{\sinh k\frac{\pi w}{2}},\quad n=\overline{1,k-1}\\
\hat{K}_{nm}(w)=\delta_{nm}-2\frac{\sinh\big[ \min(n,m)\frac{\pi w}{2}\big]\sinh\big[ (k-\max(n,m))\frac{\pi w}{2}\big]\cosh\frac{\pi w}{2}}{\sinh\big[k\frac{\pi w}{2}\big]\sinh\frac{\pi w}{2}},\quad n,m=\overline{1,k-1}\\
\hat{K}_{kk}(w)=\frac{\sinh\frac{\pi w}{2}}{\sinh \pi w}\bigg(1+\frac{\sinh[(k-1)\frac{\pi w}{2}]}{\sinh[k\frac{\pi w}{2}]}\bigg)\\
\hat{K}_{kn}=-\hat{K}_{n,k}=-e^{ -(n-k)\pi|w|/2},\quad n=\overline{k+1,\infty}\\
\hat{K}_{nm}(w)=\delta_{nm}+(e^{\pi|w|}+1)\frac{e^{-(n+m-2k)\pi|w|/2}-e^{|n-m|\pi|w|/2}}{e^{\pi|w|}-1},\quad n,m=\overline{k+1,\infty}
\end{gather*}
\begin{figure}[h]
\centering
\includegraphics[width=14cm]{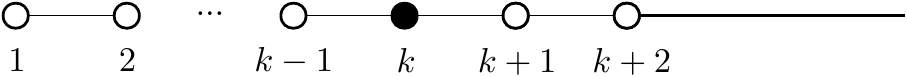}
\caption{Maximum string removed and indices rearranged}
\end{figure}
\subsection{Y system}
To transform this into the Y system, we notice that the universal kernel $s$ still satisfies the average property for the newly introduced hyperbolic kernels
\begin{align*}
\hat{s}(\hat{K}_{k,n-1}+\hat{K}_{k,n+1})=\hat{K}_{kn},\quad n=\overline{1,k-1}.
\end{align*}
As a result, deep in the left or right wing, there would be no problem. We just need to check for the three nodes $k-1,k,k+1$. We act with $-s$ on the TBA equations of $k\pm 1$
\begin{align*}
\log Y_{k-1}&=\sum_{n=1}^{k-2}K_{n,k-1}\star\log (1+Y_n)+K_{k-1,k-1}\star\log(1+Y_{k-1})+K_{k,k-1}\star\log (1+Y_k),\\
\log Y_{k+1}&=\sum_{n=k+2}^\infty K_{n,k+1}\star\log(1+Y_n)+K_{k+1,k+1}\star\log(1+Y_{k+1})+K_{k,k+1}\star\log(1+Y_k).
\end{align*}
We need the following identities
\begin{gather*}
-s\star K_{n,k-1}=K_{nk},\quad n\in\overline{1,k-2}, \quad -s\star K_{n,k+1}=K_{nk},\quad n\in \overline{k+2,\infty},\\
-s\star K_{k-1,k-1}=K_{k-1,k}-s,\quad -s\star K_{k+1,k+1}=K_{k+1,k}-s,\quad -s\star K_{k,k-1}-s\star K_{k,k+1}=K_{kk}.
\end{gather*}
Then it follows that
\begin{gather*}
-s\star\log Y_{k-1}-s\star\log Y_{k+1}=\log Y_k-s\star\log(1+Y_{k-1})-s\star\log(1+Y_{k+1}),\\
\Leftrightarrow \log Y_k=s\star\log (1+\mathcal{Y}_{k-1})+s\star(1+\mathcal{Y}_{k+1}).
\end{gather*}
The right wing is coupled to the physical node in the same way as Gross-Neveu. For the left wing, we act with s to the TBA equation of $Y_{k-2}$
\begin{align*}
\log Y_{k-2}=& K_{k,k-2}\star\log(1+Y_k)+K_{k-1,k-2}\star\log(1+Y_{k-1})+K_{k-2,k-2}\star\log(1+Y_{k-2})\\
+&\sum_{n=1}^{k-3}K_{n,k-2}\star\log(1+Y_n).
\end{align*}
With help of the following identities
\begin{gather*}
s\star K_{n,k-2}=K_{n,k-1},\quad n=\overline{1,k-3},\quad s\star K_{k,k-2}=s+K_{k,k-1},\\
s\star K_{k-1,k-2}=K_{k-1,k-1},\quad s\star K_{k-2,k-2}=s+K_{k-2,k-1}.
\end{gather*}
We obtain
\begin{gather*}
s\star\log Y_{k-2}=s\star\log(1+Y_k)+s\star\log(1+Y_{k-2})+\log Y_{k-1}\\
\Leftrightarrow \log \mathcal{Y}_{k-1}=s\star\log (1+Y_k)+s\star\log(1+\mathcal{Y}_{k-2}).
\end{gather*}
\section{Determinants}
We compute the determinants that appear in the main text. We have found these results by Mathematica. For simplicity we introduce the following notation
\begin{align*}
K_{ab}\equiv \int_{-\infty}^{+\infty}\frac{du}{2\pi}K_{ab}(u).
\end{align*}
\subsection{The IR determinant}
\label{det-IR}
We compute $\det(1-\hat{K})$ where 
\begin{align*}
\hat{K}_{ab}= K_{ab}\sqrt{\frac{Y_a^{\text{IR}}(\k)}{1+Y_a^{\text{IR}}(\k)}\frac{Y_b^{\text{IR}}(\k)}{1+Y_b^{\text{IR}}(\k)}}=[\delta_{ab}-2\min(a,b)]\frac{(1-\kappa)^2\sqrt{\kappa^a\kappa^b}}{(1-\kappa^{a+1})(1-\kappa^{b+1})},\quad  a,b\geq 1.
\end{align*}
This matrix can be implemented directly in Mathematica and we get five digit precision for twist parameters smaller than 1/2 using the first 30 magnon strings.\\
\begin{table}[!htb]
\centering
\captionsetup{justification=centering}
  \begin{tabular}{  |c | c|c|c|c| }
    \hline
    $\k=0.5$ & $\k=0.6$& $\k=0.7$ & $\k=0.8$ & $\k=0.9$ \\ \hline
   0.5 & 0.400001 & 0.30008 & 0.202126 & 0.121998 \\\hline
  \end{tabular}
  \caption{Approximation of $[\det(1-\hat{K})]^{-1}$ for some values of the twist parameter}
\end{table}

As the twist parameter tends to 1, more strings are needed to keep the precision. We can read from this numerical data that
\begin{align}
\det(1-\hat{K})=(1-\k)^{-1}.\label{IR-det}
\end{align}
This gives the loop part of IR g-function \eqref{g-GN-IR-loop}.

There is a more elegant way to obtain this result. We remark that the matrix $\hat{K}$ can be written in a slightly different way without changing the determinant of $1-\hat{K}$
\begin{align*}
\hat{K}_{ab}=[\delta_{ab}-2\min(a,b)]\frac{Y_b^{\text{IR}}(\k)}{1+Y_b^{\text{IR}}(\k)},\quad  a,b\geq 1.
\end{align*}
By factorizing the second factor we can show that \eqref{IR-det} is equivalent to
\begin{align}
\frac{\det[2Y^\text{IR}(\k)+\text{Cartan}^\text{A}_{\infty}]}{\det(\text{Cartan}^\text{A}_\infty)}=\frac{1+\k}{1-\k}.
\end{align}
From the usual method of computing the determinant of Cartan matrix of A type, we can reformulate the problem as follows. Let $G_a$ be a sequence of numbers defined by the iterative relation
    \begin{align*}
    G_{a+1}+G_{a-1}=\big[2+2Y_a^{\text{IR}}(\k)\big]G_a,\quad G_0=0,G_1=1,
    \end{align*}
then
    \begin{align}
    \lim_{a\to\infty}\frac{G_a}{a+1}=\frac{1+\k}{1-\k}.\label{Janik}
    \end{align}
We owe this derivation to Romuald Janik. Unfortunately we can only verify numerically the asymptotic \eqref{Janik}.
\subsection{The UV determinant}
\label{det-UV}
We compute $\det(1-\hat{K})$ where 
\begin{align*}
\hat{K}_{ab}=K_{ab}\sqrt{\frac{Y_a^\text{UV}(\k) Y_b^\text{UV}(\k)}{[1+Y_a^\text{UV}(\k)][1+Y_b^\text{UV}(\k)]}},
\end{align*}
with
\begin{gather*}
K_{ab}=\delta_{ab}-2\frac{\min(a,b)[k-\max(a,b)]}{k},\; a,b\in \overline{1,k-1},\quad K_{ab}=\delta_{ab}-2\min(a-k,b-k),\; a,b\geq k+1\\
K_{ka}=-K_{ak}=-\frac{a}{k},\quad 1\leq a<k,\quad K_{kk}=1-\frac{1}{2k},\quad K_{ka}=-K_{ak}=-1,\quad a\geq k+1
\end{gather*}
and 
\begin{gather*}
\frac{Y_a^\text{UV}(\k)}{1+Y_a^\text{UV}(\k)}=\k^{a}\frac{(1-\k)^2}{(1-\k^{a+1})^2}\quad a\in\overline{1,k-1}\cup\overline{k+1,\infty},\quad \frac{Y_k^\text{UV}(\k)}{1+Y_k^\text{UV}(\k)}=\frac{(1-\k^{k})(1-\k^{k+2})}{(1-\k^{k+1})^2}.
\end{gather*}
Again we choose a cut-off on $SU(2)$ magnon string length of 30. This gives five digit precision for values of the twist parameter smaller than $1/2$.\\
\begin{table}[!htb]
\centering
\captionsetup{justification=centering}
  \begin{tabular}{ |c | c | c|c|c|c| }
    \hline
     & $\k=0.5$ & $\k=0.6$& $\k=0.7$ & $\k=0.8$ & $\k=0.9$ \\ \hline
    $k=2$ & 2 & 1.6 & 1.20032 & 0.80851 & 0.487993 \\\hline
    $k=3$ & 3 & 2.40001  & 1.80048 & 1.21276& 0.731989 \\\hline
    $k=4$ & 4 & 3.20001 & 2.40064 & 1.61701& 0.975986 \\\hline
    $k=5$ & 5 & 4.00001 & 3.0008 & 2.02126& 1.21998 \\\hline
    $k=6$ & 6 & 4.80001 & 3.60096 & 2.42552&  1.46398\\\hline
    $k=7$ & 7 & 5.60002 & 4.20112 & 2.82977& 1.70797 \\ \hline
    $k=8$ & 8 & 6.40002 & 4.80128 & 3.23402& 1.95197 \\\hline
    $k=9$ & 9 & 7.20002 & 5.40144 & 3.63828& 2.19597 \\\hline
  \end{tabular}
  \caption{Approximation of $[\det(1-\hat{K})]^{-1}$ for some  twist parameters and levels}
\end{table}

We predict from this data that
\begin{align*}
\det(1-\hat{K})=[2k(1-\k)]^{-1}.
\end{align*}
This leads to the loop part of UV g-function \eqref{loop-g-level-n}.

\end{document}